\documentclass[11pt,a4paper]{article}

\usepackage{amsmath, amssymb, amsthm}
\usepackage{graphicx}
\usepackage{hyperref}
\usepackage{tocloft}
\usepackage{tikz}        
\usepackage{quantikz}    
\usepackage{multirow}

\usepackage{geometry}
\newtheorem{definition}{Definition}
\newtheorem{theorem}{Theorem}
\newtheorem{lemma}{Lemma}

\newtheorem{example}{Example}
\theoremstyle{remark}
\newtheorem{remark}{Remarque}

\begin{document}
	\title{Construction of a Family of Quantum Codes Using Sub-exceding Functions via the Hypergraph Product and the Generalized Shor Construction  }
	
	\author{Luc RABEFIHAVANANA \and Harinaivo ANDRIATAHINY \and RANDRIAMIARAMPANAHY Ferdinand}
	
	\maketitle
	
	\begin{abstract}
		In this paper, we introduce a new family of stabilizer quantum LDPC codes derived from the classical linear codes $L_k$ and $L_k^{+}$, defined via sub-exceding functions. In previous work, these codes demonstrated strong performance in minimum distance, decoding efficiency, and structural simplicity. By combining the hypergraph product framework with a generalized Shor construction, we obtain a scalable class of quantum codes with parameters $[[6k^2,\, k^2,\, d]]$. The resulting quantum codes exhibit a rich combinatorial structure and promising properties, particularly in terms of locality, low-density parity-check (LDPC) structure, and asymptotic behavior. The minimum distance satisfies $d=3$ for $k=3$ and $d=4$ for $k\ge4$, establishing a new framework for structured quantum LDPC code design and optimization.
		
		
	\end{abstract}

	\section{Introduction}
	
	\paragraph{}
	Quantum codes are essential for error correction and for ensuring the reliability of quantum computing systems. While classical computing, particularly classical binary coding, has inherent limitations, the current technological landscape is undergoing a transition toward quantum computing, which, despite its own limitations, offers significantly faster execution capabilities.
	
	In previous work, in the article entitled \emph{"Error correcting codes from sub-exceeding functions"} by L. Rabefihavanana, H. Andriatahiny, and T. Rabeherimanana, published in \emph{Computer Science Journal of Moldova, vol.27, no.1(79), 2019}, the authors introduced a classical codes $L_k$ and $L_k^+$ constructed from sub-exceeding functions. In this paper, we have a new linear error-correcting codes based on sub-exceeding functions, which are well-known combinatorial objects in permutation theory and enumerative combinatorics. The main idea of this work is to demonstrate that these functions can be exploited to construct systematic linear codes exhibiting strong properties in terms of minimum distance, algorithmic complexity, and decoding efficiency.
	
	More precisely, for any integer $k \geq 3$, the binary linear code $L_k$ has parameters $[2k, k]$, with a minimum distance equal to $3$ for $k = 3$ and $4$ for $k \geq 4$. The code $L_k^+$ has parameters $[3k, k]$, and its minimum distance is equal to $5$ for $k = 4$ and $6$ for $k \geq 5$.
	
	Now, this present work aims to extend these constructions to the quantum framework by proposing a family of quantum codes based on these two codes  $L_k$ and $L_k^+$, using the hypergraph product method and the generalized Shor construction.
	
	\section{Presentation of the codes $L_k$ and $L_k^+$}
	
	\subsection{Sub-exceeding functions}
	
	A sub-exceeding function is a mapping
	\begin{equation}
		f : \{1, 2, \ldots, k\} \longrightarrow \{0, 1, \ldots, k - 1\}
	\end{equation}
	such that
	\begin{equation}
		f(i) \leq i - 1	\text{ for all } i \in \{1, \ldots, k\}.
	\end{equation}
	
	These functions are in bijection with several classical combinatorial structures, notably Lehmer codes, permutations, and labeled trees. This combinatorial richness endows sub-exceeding functions with an algebraic structure that is particularly well suited for the construction of error-correcting codes.
	
	The authors exploit this structure to define two families of binary linear codes, denoted $L_k$ and $L_k^+$, which exhibit interesting properties in terms of parameters and performance.
	
	\subsection{The linear code $L_k$}
	
	For any integer $k \geq 3$, the code $L_k$ (\cite{Rabefihavanana2019} , \cite{Tsaratanisoa2019}) is a binary systematic linear code with parameters $L_k : [2k, k, d]$, where the length is $2k$, the dimension is $k$, and the minimum distance satisfies $d(L_k)=3$ for $k=3$ and $d(L_k)=4$ for $k\geq 4$.\\
	Thus, this code is constructed from vectors associated with sub-exceeding functions, organized systematically in the form:
	$(c \mid f(c))$,
	where $c$ is an information word and $f(c)$ is a redundancy word derived from the corresponding sub-exceeding function.\\
	As with any systematic code, its generator matrix can be written in block form as
	\begin{equation}{\label{GLk}}
		G_{L_k} = [\, I_k \mid G_k \,],
	\end{equation}
	where $I_k$ is the identity matrix of order $k$ and $G_k$ is a binary $k \times k$ matrix such that
	\begin{equation}\label{Gk}
		a_{ij} =
		\begin{cases}
			1 & \text{if } i \neq j, \\
			0 & \text{otherwise}.
		\end{cases}
	\end{equation}
	Each row of $G_k$ corresponds to the redundancy vector $f(c)$ associated with the information of $c$. Consequently, the parity-check matrix (\cite{Rabefihavanana2019}, \cite{Tsaratanisoa2019}) of $L_k$ can be written in block form as:
	\begin{equation}{\label{Hk}}
		H_{L_k} = [\, G_k \mid I_k \,].
	\end{equation}
	
	\subsection{The linear code $L_k^+$}
	For any integer $k \geq 4$, the code $L_k^+$ (\cite{Rabefihavanana2019} , \cite{Tsaratanisoa2019}) is a binary systematic linear code with parameters
	$	L_k^+ : [3k, k, d]$,where the minimum distance satisfies $d(L_k^+) = $ for $k=4$ and $d(L_k^+) = 6$ for $k\geq 4$.\\
	Its generator matrix can be written in block form as:
	\begin{equation}{\label{GLk+}}
		G_{L_k^+} = [\, I_k \mid G_k \mid I_k \,],
	\end{equation}
	where $I_k$ is the identity matrix of order $k$ and $G_k$ is a binary $k \times k$ matrix defined in \ref{Gk}.\\
	Moreover, the parity-check matrix (\cite{Rabefihavanana2019} , \cite{Tsaratanisoa2019}) of $L_k^+$ can be written in block form as:
	\begin{equation}{\label{HLk+}}
		H_{L_k^+} =
		\begin{pmatrix}
			G_k & I_k & 0_k \\
			I_k & 0_k & I_k
		\end{pmatrix}.
	\end{equation}
	
	\subsection{Explicit examples for the two codes $L_k$ and $L_k^{+}$}
	For the code $L_k$, let us consider $k=4$. We obtain:
	\begin{equation*}
		G_{\mathcal{L}_{4}}= \left(\begin{array}{cccccccc}
			1 & 0 & 0 & 0 & 0 & 1 & 1 & 1\\
			0 & 1 & 0 & 0 & 1 & 0 & 1 & 1\\
			0 & 0 & 1 & 0 & 1 & 1 & 0 & 1\\
			0 & 0 & 0 & 1 & 1 & 1 & 1 & 0\\
		\end{array}\right) 
	\end{equation*}
	and the codewords are: 
	\begin{equation*}
		\mathcal{L}_{4}=\begin{array}{c}
			0000 0000 \\
			0001 1110 \\
			0010 1101 \\
			0100 1011 \\
			1000 0111 \\
		\end{array}
		\begin{array}{c}
			0011 0011 \\
			0101 0101 \\
			1001 1001 \\
			0110 0110 \\
			1010 1010 \\
			1100 1100 \\
		\end{array}
		\begin{array}{c}
			0111 1000 \\
			1011 0100 \\
			1101 0010 \\
			1110 0001 \\
			1111 1111
		\end{array}
	\end{equation*} 
	For the code $L_k^{+}$, let us consider the example $k=5$. Thus,
	\begin{equation*}
		G_{\mathcal{L}^{+}_{5}}=  \left(\begin{array}{ccccccccccccccc}
			1 & 0 & 0 & 0 & 0 & 0 & 1 & 1 & 1 & 1 &  1 & 0 & 0 & 0 & 0\\
			0 & 1 & 0 & 0 & 0 & 1 & 0 & 1 & 1 & 1 & 0 & 1 & 0 & 0 & 0\\
			0 & 0 & 1 & 0 & 0 & 1 & 1 & 0 & 1 & 1 & 0 & 0 & 1 & 0 & 0\\
			0 & 0 & 0 & 1 & 0 & 1 & 1 & 1 & 0 & 1 & 0 & 0 & 0 & 1 & 0\\
			0 & 0 & 0 & 0 & 1 & 1 & 1 & 1 & 1 & 0 &  0 & 0 & 0 & 0 & 1\\
		\end{array}\right). 
	\end{equation*}
	Regarding the parity-check matrices:
	\begin{itemize}
		\item[-] for $L_4$, we have:
		\begin{equation*}
			H_{\mathcal{L}_{4}}= \left(\begin{array}{cccccccc}
				0 & 1 & 1 & 1 & 1 & 0 & 0 & 0 \\
				1 & 0 & 1 & 1 & 0 & 1 & 0 & 0 \\
				1 & 1 & 0 & 1 & 0 & 0 & 1 & 0 \\
				1 & 1 & 1 & 0 & 0 & 0 & 0 & 1 \\
			\end{array}\right)
		\end{equation*}
		
		\item[-] for $L_4^{+}$, the parity-check matrix is
		\begin{equation*}
			H_{(\mathcal{L}_{4}^{+})}=   \left(\begin{array}{cccccccccccc}
				0 & 1 & 1 & 1 & 1 & 0 & 0 & 0 & 0 & 0 & 0 & 0 \\
				1 & 0 & 1 & 1 & 0 & 1 & 0 & 0 & 0 & 0 & 0 & 0 \\
				1 & 1 & 0 & 1 & 0 & 0 & 1 & 0 & 0 & 0 & 0 & 0  \\
				1 & 1 & 1 & 0 & 0 & 0 & 0 & 1 & 0 & 0 & 0 & 0 \\
				1 & 0 & 0 & 0 & 0 & 0 & 0 & 0 & 1 & 0 & 0 & 0 \\
				0 & 1 & 0 & 0 & 0 & 0 & 0 & 0 & 0 & 1 & 0 & 0 \\
				0 & 0 & 1 & 0 & 0 & 0 & 0 & 0 & 0 & 0 & 1 & 0 \\
				0 & 0 & 0 & 1 & 0 & 0 & 0 & 0 & 0 & 0 & 0 & 1
			\end{array}\right).
		\end{equation*}
	\end{itemize}
	
	\section{Fundamentals and Structure of Stabilizer Codes}
	
	\subsection{Some quantum notions}
	
	\begin{definition}[Hilbert space of a qubit \cite{Tapp1999}, \cite{Gilyen2019}, \cite{Steane1996},\cite{Bodin2024}]
		The Hilbert space of a qubit is defined as
		\begin{equation}
			\mathcal{H} =
			\left\{
			\alpha |0\rangle + \beta |1\rangle
			\;\middle|\;
			\alpha,\beta \in \mathbb C
			\right\},
		\end{equation}
		where $|0\rangle$ and $|1\rangle$ form an orthonormal basis, that is,
		$$
		\langle 0 | 1 \rangle = 0.
		$$
		
		A qubit is a state $|\psi\rangle \in \mathcal H$ such that
		\begin{equation}
			|\psi\rangle = \alpha |0\rangle + \beta |1\rangle,
			\qquad
			|\alpha|^2 + |\beta|^2 = 1.
		\end{equation}
	\end{definition}
	
	\begin{definition}[Tensor space of $n$ qubits \cite{Tapp1999}, \cite{Steane1996}, \cite{Baboin2011},\cite{Bodin2024}]
		The Hilbert space of a register of $n$ qubits is defined by
		\begin{equation}
			\mathcal H^{\otimes n}=
			\underbrace{\mathcal H \otimes \mathcal H \otimes \cdots \otimes \mathcal H}_{n}
			=
			\left\{
			\sum_{x \in \{0,1\}^n} \alpha_x |x\rangle
			\;\middle|\;
			\alpha_x \in \mathbb C
			\right\}.
		\end{equation}
		An $n$-qubit register is a state $|\psi\rangle \in \mathcal H^{\otimes n}$ such that
		\begin{equation}
			|\psi\rangle = \sum_{x \in \{0,1\}^n} \alpha_x |x\rangle,
			\qquad
			\sum_{x \in \{0,1\}^n} |\alpha_x|^2 = 1.
		\end{equation}
	\end{definition}
	
	\begin{definition}[Quantum code \cite{Steane1996},\cite{Bodin2024}, \cite{Calderbank1997}]
		A quantum code with parameters $[[n,k]]$ is a subspace	$
		\mathcal C \subset \mathcal H^{\otimes n}
		$
		of dimension $2^k$, used to encode $k$ logical qubits
		into $n$ physical qubits.
		The rate of the code is defined as $\frac{k}{n}$ and, more generally, as
		\begin{equation}
			R = \frac{\log_2 (\dim C)}{n}.
		\end{equation}
	\end{definition}
	
	\subsection{Pauli groups and stabilizer codes}
	
	\begin{definition}
		Recall the Pauli matrices, which form an orthonormal basis
		of the space of $2\times2$ matrices:
		\begin{equation}
			I=\begin{pmatrix}1&0\\0&1\end{pmatrix},\;
			X=\begin{pmatrix}0&1\\1&0\end{pmatrix},\;
			Y=\begin{pmatrix}0&-i\\i&0\end{pmatrix},\;
			Z=\begin{pmatrix}1&0\\0&-1\end{pmatrix}.
		\end{equation}
	\end{definition}
	
	Elementary errors acting on a qubit are described by these Pauli operators.
	These matrices satisfy the following relations:
	\begin{equation}
		\left\lbrace
		\begin{array}{c}
			X^{2}=Y^{2}=Z^{2}=I,\\
			XY =-YX = iZ,\\
			YZ = -ZY = -iX,\\
			ZX = -XZ = iY.
		\end{array}
		\right.
	\end{equation}
	
	Two Pauli matrices $E_1$ and $E_2$ commute if $E_1E_2=E_2E_1$ and anticommute if $E_1E_2=-E_2E_1$. In particular, a Pauli matrix commutes with itself and with $I$, and anticommutes with the two other nontrivial Pauli matrices.
	
	\begin{definition}[Pauli group for $n$ qubits \cite{Tapp1999},\cite{Steane1996},\cite{Tillich2008},\cite{Delfosse2012} ]
		The $n$-qubit Pauli group is defined as
		\begin{equation}
			\mathcal{P}_n = \left\lbrace  1,-1,i,-i \right\rbrace \times \left\lbrace I,X,Y,Z \right\rbrace ^{\otimes n}.
		\end{equation}
	\end{definition}
	
	The Pauli group plays a fundamental role in the construction of $n$-qubit stabilizer codes.
	
	\begin{definition}[Stabilizer group \cite{Steane1998}, \cite{Calderbank1997}, \cite{Lai2011}, \cite{Cohen1998} ]
		A stabilizer group is a commutative subgroup $S$ of $\mathcal{P}_n$
		that does not contain $-I$. The stabilizer code $C(S)$ associated with $S$
		is the set of fixed points of $S$ in $\mathcal{H}^{\otimes n}$:
		
		\begin{equation}
			C(S) = \{\, |\psi\rangle \in \mathcal{H}^{\otimes n} \text{ such that } s |\psi\rangle = |\psi\rangle, \; \forall s \in S \,\}.
		\end{equation}
		The integer $n$ is the length of the quantum code.
	\end{definition}
	
	Assume that $S = \langle S_1, S_2, \dots, S_r \rangle$ is generated by $r$ generators $S_i \in \mathcal{P}_n$.
	
	\begin{definition}
		The stabilizer matrix of $C(S)$ is the matrix $H \in M_{r,n}(\{I,X,Y,Z\})$ whose $i$-th row represents the generator $S_i$.  
		The entry $H_{i,j}$ is the $j$-th component of $S_i$.
	\end{definition}
	
	A stabilizer code is completely specified by its stabilizer matrix, although several different matrices may define the same stabilizer group and the same code. This matrix can be viewed as the quantum analogue of the parity-check matrix of a classical code.
	
	\begin{theorem}[Parameters of stabilizer codes \cite{Lai2011}, \cite{Cohen1998} , \cite{Delfosse2012}]
		A $[[n,k,d]]$ quantum stabilizer code $\mathcal{C}(\mathcal{S})$ is a subspace of $\mathcal{H}^{\otimes n}$ such that:
		\begin{itemize}
			\item $\mathcal{C}(\mathcal{S})$ has dimension $2^k$ and possesses $r=n-k$ independent generators,
			\item The minimum distance $d$ is the minimum weight of a nontrivial undetectable error:
			\begin{equation}
				d = \min \{ w(E) \mid E \in P_n \setminus S, \; \sigma(E) = 0 \},
			\end{equation}
			where $w(E)$ denotes the weight of $E$, i.e., the number of components different from the identity.
		\end{itemize}
	\end{theorem}
	
	\subsection{Symplectic product and construction of a stabilizer code from two classical binary codes}
	
	An element of $\mathcal{G}_n$ can be decomposed as
	$i^{c}\,M_{1}\otimes M_{2}\otimes \cdots \otimes M_{n}$,
	where each $M_j$ is a Pauli operator $I,X,Y,$ or $Z$, and
	$c\in \{0,1,2,3\}$. For $E\in \mathcal{G}_n$, we denote
	$i^c=\lambda$ and write
	\begin{equation}
		E=\lambda\bigotimes_{i=1}^{n} E(i).
	\end{equation}
	\begin{definition}[Symplectic representation \cite{Lai2011} , \cite{Delfosse2012}]
		Let $g=i^{c}M_{1}\otimes M_{2}\otimes\cdots\otimes M_{n}$ be an element of $\mathcal{G}_n$. The symplectic representation is the mapping
		$$
		\begin{array}{c l c l}
			\varphi:& \mathcal{G}_n & \longrightarrow & \mathbb{Z}_2^{2n}\\
			& g & \longmapsto & (\alpha_1,\cdots,\alpha_n \mid \beta_1,\cdots,\beta_n)
		\end{array}
		$$
		where for all $i\in \{1,\ldots,n\}$,
		\begin{equation}
			\left\{
			\begin{array}{ccc}
				\alpha_i=0 & \text{if} & M_i\in  \{I,Z\},\\
				\alpha_i=1 & \text{if} & M_i\in \{X,Y\},\\
				\beta_i = 0 & \text{if}  & M_i \in \{I,X\},\\
				\beta_i = 1 & \text{if}  & M_i \in \{Y,Z\}.
			\end{array}
			\right.
		\end{equation}
	\end{definition}
	
	\begin{example}\label{exemple_matrice_forme_symplectique}
		Consider the three Pauli operators defining the matrix $H$ of a quantum code acting on $n=4$ qubits:
		$$
		H=\begin{pmatrix}
			Z&Z&I&I\\
			I&Z&Z&I\\
			I&I&Z&Z
		\end{pmatrix}.
		$$
		The three generators are
		$g_1=Z\otimes Z\otimes I\otimes I$, 
		$g_2=I\otimes Z\otimes Z\otimes I$ and 
		$g_3=I\otimes I\otimes Z\otimes Z$.
		Their symplectic representations are
		\begin{itemize}
			\item [-] $\varphi(g_1)=(0,\,0,\,0,\,0\mid 1,\,1,\,0,\,0)$,
			\item [-] $\varphi(g_2)=(0,\,0,\,0,\,0\mid 0,\,1,\,1,\,0)$,
			\item [-] $\varphi(g_3)=(0,\,0,\,0,\,0\mid 0,\,0,\,1,\,1)$.
		\end{itemize}
		The symplectic form of $H$ is therefore
		$$
		H_{\text{symplectic}}=\left[ H_X \;\middle|\; H_Z \right]
		=
		\left[
		\begin{array}{cccc}
			0&0&0&0\\
			0&0&0&0\\
			0&0&0&0
		\end{array}
		\;\middle\|\;
		\begin{array}{cccc}
			1&1&0&0\\
			0&1&1&0\\
			0&0&1&1
		\end{array}
		\right].
		$$
	\end{example}
	
	\begin{definition}
		Let $\{g_1, g_2,\ldots ,g_r\}$ be a set of $r$ independent generators of a stabilizer group $\mathcal{S}$.  
		We define the \emph{parity-check matrix} $H$ of $\mathcal{S}$ such that the $i$-th row of $H$ is $\varphi(g_i)$.  
		
		Then $H$ is a binary matrix of size $r\times 2n$, and we write
		\begin{equation}
			H=[H_X \mid H_Z],
		\end{equation}
		where $H_X$ and $H_Z$ are binary matrices of size $r\times n$.
	\end{definition}
	
	Since the stabilizer group $\mathcal{S}$ is abelian, we have $gh=hg$ for all $g,h\in \mathcal{S}$. Using the homomorphism $\varphi$, this implies
	\begin{equation}
		\varphi(g)\Lambda_{2n}\varphi(h)^T=0,
		\quad \forall g,h\in \mathcal{S},
		\label{equation_commutativité}
	\end{equation}
	where
	$$
	\Lambda_{2n}=
	\begin{bmatrix}
		0_{n\times n}&I_{n\times n}\\
		I_{n\times n}&0_{n\times n}
	\end{bmatrix}.
	$$
	
	\begin{lemma}[Commutation condition \cite{Delfosse2012}]\label{Condition de la commutativité}
		The parity-check matrix $H$ of a stabilizer group $\mathcal{S}$ must satisfy
		\begin{equation}
			H\Lambda_{2n}H^{T}
			=
			H_{X}H^{T}_{Z}+H_{Z}H^{T}_{X}
			=
			0_{r\times r},
		\end{equation}
		where $0_{i\times j}$ denotes the zero matrix of size $i\times j$.
	\end{lemma}
	
	Starting from two classical binary codes defined by their parity-check matrices, we can construct a quantum stabilizer code. The key condition is that the symplectic product between the two matrices must vanish, ensuring that the corresponding quantum generators commute. Using these matrices, one defines $X$-type and $Z$-type generators that form the stabilizer group of the quantum code.
	
	\subsection{Quantum code construction via the hypergraph product method}
	
	Let $C_1$ and $C_2$ be respectively two classical linear codes with parameters $[n_1,k_1,d_1]$ and $[n_2,k_2,d_2]$, and parity-check matrices $H_1$ and $H_2$. Define the matrices $H_X$ and $H_Z$ (\cite{Pradhan2025}, \cite{Tillich2008}) by
	\begin{equation}{\label{HxHz_hyperb}}
		H_X = \begin{pmatrix}H_1 \otimes I_{n_2}  & I_{r_1}\otimes H^{T}_2 \end{pmatrix}
		\;\text{and}\;
		H_Z = \begin{pmatrix}I_{n_1} \otimes H_2 & H^{T}_1\otimes I_{r_2} \end{pmatrix},
	\end{equation}
	where $I_i$ denotes the identity matrix of size $i\times i$, $\otimes$ is the tensor (Kronecker) product and $r_i$ is the integer such that $r_i=n_i - k_i$.\\
	The parity-check matrix \cite{Pradhan2025} of the resulting quantum code $\mathcal{Q}$ is defined by
	\begin{equation}{\label{Qmatrix_shor}}
		H=\begin{pmatrix}H_X && 0\\0 && H_Z\end{pmatrix}.
	\end{equation}
	In this construction:
	\begin{itemize}
		\item[-]$H_Z$ (Z-type checks) combines the parity constraints of $C_1$ expanded over $C_2$ (first block) with the parity constraints of $C_2$ replicated along the parity rows of $C_1$ (second block).
		\item[-]$H_X$ (X-type checks) follows a similar construction using dual structures of the classical codes.
	\end{itemize}
	
	The parameters (\cite{Pradhan2025}, \cite{Tillich2008})  of the resulting quantum code $\mathcal{Q}$ are denoted by $[[n,k,d]]$ and given by
	\begin{equation}{\label{parametre_hyperb}}
		[[n,k,d]]=[[n_1n_2+r_1r_2,\, k_1k_2+k^{\bot}_1k^{\bot}_2,\, \min(d_1,d_2,d^{\bot}_1,d^{\bot}_2)]].
	\end{equation}
	\begin{itemize}
		\item[-] $n$ is the total number of qubits, given by $n=n_1n_2+r_1r_2$,
		\item[-] $k$ is the logical dimension (number of encoded logical qubits),
		$$
		k=k_1k_2+k^{\bot}_1k^{\bot}_2,
		$$
		\item[-] $d$ is the minimum distance of the resulting code, related to the distances of the initial codes by
		$$
		d= \min(d_1,d_2,d^{\bot}_1,d^{\bot}_2).
		$$
	\end{itemize}
	The hypergraph product construction satisfies the necessary CSS condition since the symplectic product vanishes ($H_XH^{T}_Z=0$):
	\begin{equation*}
		\begin{array}{ccc}
			H_X.H^{T}_Z &=& \begin{pmatrix}H_1 \otimes I_{n_2} & & I_{r_1}\otimes H^{T}_2\end{pmatrix}
			\begin{pmatrix}I_{n_1} \otimes H_2 && H^{T}_1\otimes I_{r_2} \end{pmatrix}^{T}\\
			&=&\begin{pmatrix}H_1 \otimes I_{n_2}  && I_{r_1}\otimes H^{T}_2\end{pmatrix}
			\begin{pmatrix}(I_{n_1} \otimes H_2)^{T} \\ (H^{T}_1\otimes I_{r_2})^{T} \end{pmatrix}\\
			&=&(H_1\otimes I_{n_2})(I_{n_1}\otimes H_2)^{T}
			+ (I_{r_1}\otimes H^{T}_2)(H^{T}_1\otimes I_{r_2})^{T}\\
			&=&(H_1\otimes I_{n_2})(I_{n_1}\otimes H_2^{T})
			+ (I_{r_1}\otimes H^{T}_2)(H_1\otimes I_{r_2})\\
		\end{array}.
	\end{equation*}
	Since $(A\otimes B)(C\otimes D)=AC\otimes BD$,
	\begin{equation*}
		\begin{array}{ccc}
			H_X.H^{T}_Z
			&=&H_1I_{n_1}\otimes I_{n_2}H^{T}_2
			+ I_{r_1}H_1\otimes H^{T}_2 I_{r_2}\\
			&=&H_1\otimes H^{T}_2 + H_1 \otimes H^{T}_2\\
			&=&0.
		\end{array}
	\end{equation*}
	
	However, constructing a quantum LDPC code using this method may be delicate due to orthogonality constraints. Nevertheless, if both classical codes $C_1$ and $C_2$ are LDPC codes, a QLDPC construction can be achieved.
	
	\subsection{Generalized Shor construction method}
	The generalized Shor construction is similar to the hypergraph product construction. The main difference lies in the structure of the matrices $H_X$ and $H_Z$, defined as
	\begin{equation}{\label{HxHz_shor}}
		H_X = H_1 \otimes I_{n_2}
		\qquad\text{and}\qquad
		H_Z = G_1 \otimes H_2,
	\end{equation}
	where $G_1$ is the generator matrix of $C_1$. The resulting quantum code $\mathcal{Q}$ obtained from the classical linear codes $C_1$ and $C_2$ has parameters (\cite{Pradhan2025}, \cite{Tillich2008})
	\begin{equation}{\label{parametre_shor}}
		[[n,k,d]]=[[n_1n_2,\, k_1k_2,\, \min(d_1,d_2)]].
	\end{equation}
	This generalized Shor construction produces a CSS code associated with the pair $(C_X,\,C_Z)$ satisfying $C^{\perp}_X \subseteq C_Z$. This relation can be verified by computing
	\begin{align*}
		H_X.H^{T}_Z
		&=(H_1 \otimes I_{n_2})(G_1 \otimes H_2)^{T}\\
		&=(H_1 \otimes I_{n_2})(G^{T}_1 \otimes H^{T}_2)\\
		&=(H_1G^{T}_1) \otimes (I_{n_2}H^{T}_2).
	\end{align*}
	Since $G_1$ and $H_1$ are respectively the generator and parity-check matrices of $C_1$, we have $H_1G^{T}_1=0$. Therefore,
	\begin{equation}
		H_X.H^{T}_Z=[0\otimes (I_{n_2}H^{T}_2)] = 0.
	\end{equation}
	\section{Quantum code obtained from the two codes $\mathcal{L}_k$ and $\mathcal{L}^{+}_k$}
	
	Consider two codes $\mathcal{L}_{k}$ and $\mathcal{L}^{+}_{k'}$ with respective parameters $[2k,\,k,\,d]$ and $[3k',\,k',\,d']$. Let $G_{\mathcal{L}_{k}}$ and $G_{\mathcal{L}^{+}_{k'}}$ denote their respective generator matrices defined as above. Using $G_{\mathcal{L}_{k}}$ and $G_{\mathcal{L}^{+}_{k'}}$, we can determine their corresponding parity-check matrices $H_{\mathcal{L}_{k}}$ and $H_{\mathcal{L}^{+}_{k'}}$.
	
	\begin{itemize}
		\item[-] For $G_{\mathcal{L}_{k}}$, set $-A^{T}_{\mathcal{L}_{k}}=G_{k}$ \ref{GLk}. The generator matrix then becomes
		$
		G_{\mathcal{L}_{k}}=\begin{pmatrix}I_k & -A^{T}_{\mathcal{L}_{k}}\end{pmatrix}
		$
		of dimension $k\times 2k$. Thus,
		$
		H_{\mathcal{L}_{k}} = \begin{pmatrix}A_{\mathcal{L}_{k}} & I_{2k-k}\end{pmatrix}
		= \begin{pmatrix}A_{\mathcal{L}_{k}} & I_{k}\end{pmatrix},
		$
		where $H_{\mathcal{L}_{k}}$ has dimension $k\times 2k$.
		
		\item[-] For $G_{\mathcal{L}^{+}_{k}}$, set
		$
		-A^{T}_{\mathcal{L}^{+}_{k}}=\begin{pmatrix}G_k & I_k\end{pmatrix}
		$\ref{GLk+}.
		The generator matrix becomes
		$
		G_{\mathcal{L}^{+}_{k}}=\begin{pmatrix}I_{k} & -A^{T}_{\mathcal{L}^{+}_{k}}\end{pmatrix}
		$
		of dimension $k\times 3k$. Hence,
		$
		H_{\mathcal{L}^{+}_{k}}=\begin{pmatrix}A_{\mathcal{L}^{+}_{k}} & I_{2k}\end{pmatrix},
		$
		where $H_{\mathcal{L}^{+}_{k}}$ has dimension $2k\times 3k$.
	\end{itemize}
	
	We now apply construction via the hypergraph product \ref{HxHz_hyperb} and the generalized Shor \ref{HxHz_shor}  method to build the quantum code derived from $\mathcal{L}_{k}$ and $\mathcal{L}^{+}_{k}$, which we denote respectively by $C_1$ and $C_2$.
	\begin{equation}
		H_X=H_{\mathcal{L}_{k}}\otimes I_{3k}
		=\begin{pmatrix}A_{\mathcal{L}_{k}} && I_{k}\end{pmatrix}\otimes I_{3k}
	\end{equation}
	and
	\begin{equation}
		H_Z=G_{\mathcal{L}_{k}}\otimes H_{\mathcal{L}^{+}_{k}} =\begin{pmatrix}I_k && -A^{T}_{\mathcal{L}_{k}}\end{pmatrix}\otimes
		\begin{pmatrix}A_{\mathcal{L}^{+}_{k}} && I_{2k}\end{pmatrix}.
	\end{equation}
	The parity-check matrix $H$ of the resulting CSS quantum code  \ref{Qmatrix_shor} is therefore
	\begin{equation}\label{controle_CSS}
		H=\begin{pmatrix}
			\begin{pmatrix}A_{\mathcal{L}_{k}} && I_{k}\end{pmatrix}\otimes I_{3k} && 0 \\
			0 && \begin{pmatrix}I_k && -A^{T}_{\mathcal{L}_{k}}\end{pmatrix}\otimes
			\begin{pmatrix}A_{\mathcal{L}^{+}_{k}} && I_{2k}\end{pmatrix}
		\end{pmatrix}.
	\end{equation}
	
	The parameters (\ref{parametre_hyperb} ,  \ref{parametre_shor}) of this quantum code are
	\begin{equation}
		[[n_1n_2,\,k_1k_2,\,\min(d_1,\,d_2)]]=[[6k^{2},\,k^{2},\,\min(d,\,d')]].
	\end{equation}
	Collecting the results:
	
	\begin{itemize}
		\item[-] For every positive integer $k$, the code $L_k$ has parameters $[2k, k]$. Its minimum distance is equal to $3$ for $k = 3$ and $4$ for $k \geq 4$.
		\item[-] The code $L_{k}^{+}$ has parameters $[3k, k]$. Its minimum distance is equal to $5$ for $k = 4$ and $6$ for $k \geq 5$.
	\end{itemize}
	
	Therefore, for every positive integer $k$, we obtain a family of quantum codes $QL_k$ with parameters
	\begin{equation}
		\mathcal{QL}_k=[[6k^2, k^2, d]],
	\end{equation}
	where the minimum distance of this quantum LDPC code equals $3$ for $k = 3$ and $4$ for $k \geq 4$.
	
	\subsection{Rich combinatorial structure}
	The present constructed family of quantum codes relies on a nontrivial interaction between two families of classical linear codes, namely $L_k$ and $L_k^+$, The orthogonality relations, the systematically organized parity-check matrices $H_X$ and $H_Z$, give rise to highly structured Tanner graphs. Such regularity not only facilitates the analysis of code parameters but also enables natural generalizations of the construction.
	\subsection{Minimum distance and error robustness}
	The minimum distance of the resulting quantum code family is directly inherited from the properties of the underlying classical codes. The fact that the distance is strictly greater than 2, and in particular reaches the value $4$ for all $k \geq 4$, guarantees the ability to detect any low-weight error and to correct at least one arbitrary quantum error.\\
	Although the distance remains bounded, this behavior is consistent with many known quantum LDPC constructions, where locality and efficient decodability are prioritized over rapidly growing distance. The stability of the distance for $k \geq 4$ nevertheless reflects a non-accidental structural robustness of the construction.
	\subsection{Asymptotic parameters and code rate}
	From an asymptotic perspective, the quantum code family has parameters $[6k^2, k^2, d]$, which implies a constant rate equal to $1/6$ as $k \to \infty$. Achieving simultaneously a constant rate and an LDPC structure constitutes a notable result, given the strong commutation constraints imposed by stabilizer quantum codes.\\
	This property places the construction within a restricted class of nonzero-rate quantum LDPC codes, making it a relevant candidate for large-scale quantum error-correction schemes.
	\section{Practical Examples}
	\begin{example}
		In this example, we construct a quantum code from $\mathcal{L}_4$ and $\mathcal{L}^{+}_4$, which are classical codes with parameters $[8,4,4]$ and $[12,4,5]$, respectively.\\
		For $\mathcal{L}_4$, we have $G_{\mathcal{L}_4}=\begin{pmatrix}I_4 & G_4\end{pmatrix}$ where 
		$G_4=
		\left( \begin{array}{cccc}
			0 & 1 & 1 & 1 \\
			1 & 0 & 1 & 1 \\
			1 & 1 & 0 & 1 \\
			1 & 1 & 1 & 0
		\end{array}\right) .
		$
		Consequently, the parity-check matrix is $H_{\mathcal{L}_{4}} = \begin{pmatrix} G_4 & I_4 \end{pmatrix},$
		which has dimensions $4 \times 8$.\\
		For $\mathcal{L}^{+}_4$, the generator matrix is initially $G_{\mathcal{L}^{+}_4} = \begin{pmatrix} I_4 & G_4 & I_4 \end{pmatrix}.$ By setting $ -A^{T}_{\mathcal{L}^{+}_4} = \begin{pmatrix} G_4 & I_4 \end{pmatrix},$ it can equivalently be written as $G_{\mathcal{L}^{+}_4} = \begin{pmatrix} I_4 & -A^{T}_{\mathcal{L}^{+}_4} \end{pmatrix},$ and the corresponding parity-check matrix becomes $H_{\mathcal{L}^{+}_4} = \begin{pmatrix} A_{\mathcal{L}^{+}_4} & I_8 \end{pmatrix},$ with dimensions $8 \times 12$.\\
		\begin{equation*}
			H_{\mathcal{L}^{+}_{4}} =
			\left( \begin{array}{cccccccccccc}
				0 & 1 & 1 & 1 & 1 & 0 & 0 & 0 & 0 & 0 & 0 & 0\\
				1 & 0 & 1 & 1 & 0 & 1 & 0 & 0 & 0 & 0 & 0 & 0\\
				1 & 1 & 0 & 1 & 0 & 0 & 1 & 0 & 0 & 0 & 0 & 0\\
				1 & 1 & 1 & 0 & 0 & 0 & 0 & 1 & 0 & 0 & 0 & 0\\
				1 & 0 & 0 & 0 & 0 & 0 & 0 & 0 & 1 & 0 & 0 & 0\\
				0 & 1 & 0 & 0 & 0 & 0 & 0 & 0 & 0 & 1 & 0 & 0\\
				0 & 0 & 1 & 0 & 0 & 0 & 0 & 0 & 0 & 0 & 1 & 0\\
				0 & 0 & 0 & 1 & 0 & 0 & 0 & 0 & 0 & 0 & 0 & 1
			\end{array}\right).
		\end{equation*}
		
		We now apply the hypergraph product method and the generalized Shor construction to build the quantum code from $\mathcal{L}_{4}$ and $\mathcal{L}^{+}_{4}$, denoted $C_1$ and $C_2$ respectively.\\
		The parity-check matrix $H$ of the resulting  quantum code is defined, according to equation (\ref{HxHz_hyperb} , \ref{HxHz_shor}), by
		\begin{align*}
			H &= \begin{pmatrix} H_X & 0 \\ 0 & H_Z \end{pmatrix} \\
			&= \begin{pmatrix}
				\begin{pmatrix}A_{\mathcal{L}_{4}} & I_4\end{pmatrix}\otimes I_{12} & 0 \\
				0 & \begin{pmatrix}I_4 & -A^{T}_{\mathcal{L}_{4}}\end{pmatrix}\otimes \begin{pmatrix}A_{\mathcal{L}^{+}_{4}} & I_8\end{pmatrix}
			\end{pmatrix}.
		\end{align*}
		The resulting quantum code has parameters:
		$$
		\mathcal{QL}_4= [[n_1 n_2,\, k_1 k_2,\, \min(d_1,d_2)]] = [[96, 16, 4]].
		$$
	\end{example}
	
	\begin{example}
		In this second example, we construct a quantum code from $\mathcal{L}_5$ and $\mathcal{L}^{+}_5$, which are classical codes with parameters $[10,5,4]$ and $[15,5,6]$, respectively.\\
		Now, we have $G_{\mathcal{L}_5}=\begin{pmatrix}I_5 & G_5\end{pmatrix}$ where 
		$G_5=
		\left( \begin{array}{ccccc}
			0 & 1 & 1 & 1 & 1\\
			1 & 0 & 1 & 1 & 1\\
			1 & 1 & 0 & 1 & 1\\
			1 & 1 & 1 & 0 & 1\\
			1 & 1 & 1 & 1 & 0
		\end{array}\right) .
		$
		Consequently, the parity-check matrix is $H_{\mathcal{L}_{5}} = \begin{pmatrix} G_5 & I_5 \end{pmatrix},$
		which has dimensions $5 \times 10$.\\
		For $\mathcal{L}^{+}_5$, the generator matrix is initially $G_{\mathcal{L}^{+}_5} = \begin{pmatrix} I_5 & G_5 & I_5 \end{pmatrix}.$ By setting $ -A^{T}_{\mathcal{L}^{+}_5} = \begin{pmatrix} G_5 & I_5 \end{pmatrix},$ it can equivalently be written as $G_{\mathcal{L}^{+}_5} = \begin{pmatrix} I_5 & -A^{T}_{\mathcal{L}^{+}_5} \end{pmatrix},$ and the corresponding parity-check matrix becomes $H_{\mathcal{L}^{+}_5} = \begin{pmatrix} A_{\mathcal{L}^{+}_5} & I_{10} \end{pmatrix},$ with dimensions $10 \times 15$.\\
		The parity-check matrix $H$ of the resulting  quantum code is defined, according to equation \ref{HxHz_hyperb} , \ref{HxHz_shor}, by
		\begin{align*}
			H_X &= H_{\mathcal{L}_5} \otimes I_{15}, \\
			H_Z &= G_{\mathcal{L}_5} \otimes H_{\mathcal{L}^{+}_5}, \\
			H &= \begin{pmatrix} H_X & 0 \\ 0 & H_Z \end{pmatrix} \\
			&= \begin{pmatrix}
				\begin{pmatrix} A_{\mathcal{L}_5} & I_5 \end{pmatrix} \otimes I_{15} & 0 \\
				0 & \begin{pmatrix} I_5 & -A^{T}_{\mathcal{L}_5} \end{pmatrix} \otimes \begin{pmatrix} A_{\mathcal{L}^{+}_5} & I_{10} \end{pmatrix}
			\end{pmatrix}.
		\end{align*}
		
		The resulting quantum code has parameters:
		$$
		\mathcal{QL}_5 = [[n_1 n_2, k_1 k_2, \min(d_1,d_2)]] = [[120, 25, 4]].
		$$
	\end{example}

	\section{Encoding and Decoding Procedures for the new quantum code $\mathcal{QL}_k$}
	
	In this section, we provide a detailed description of the encoding and decoding procedures associated with the new quantum codes constructed  $\mathcal{QL}_k$. The goal is to establish an explicit operational framework clarifying how logical information is embedded into the stabilizer subspace and how error correction can be implemented efficiently.
	
	\subsection{CSS Encoding Structure}
	
	The quantum code $\mathcal{QL}_k$ arises from a CSS construction defined by two binary linear codes through the stabilizer matrices
	$$H_X = H_{L_k} \otimes I_{3k}, \qquad
	H_Z = G_{L_k} \otimes H_{L_k^+}.
	$$
	These matrices define respectively the $X$-type and $Z$-type stabilizer generators. The commutation condition
	$H_X H_Z^T = 0
	$ ensures that the resulting stabilizer group defines a valid quantum CSS code.
	\subsection{Initialization of the Register}
	
	The encoding procedure begins with the preparation of an $6k^{2}$-qubit register:
	
	\begin{itemize}
		
		\item [-] The first $k^{2}$ qubits contain the logical input state
		\[
		|\psi_L\rangle = \sum_{x \in \{0,1\}^{k^{2}}} \alpha_x |x\rangle.
		\]
		
		\item [-] The remaining $6k^{2}-k^{2}$ qubits are initialized in the computational basis state $|0\rangle$ and serve as redundancy and parity qubits.
		
	\end{itemize}
	
	
	\subsection{Quantum encoding of the newly constructed code}
	
	From the construction of the matrices $H_X$ and $H_Z$, defined as
	\[
	H_X = H_{L_k}\otimes I_{3k}
	\qquad \text{and} \qquad
	H_Z = G_{L_k}\otimes H_{L_k^{+}},
	\]
	we obtain two matrices of respective dimensions
	$3k^{2}\times 6k^{2}$ and $2k^{2}\times 6k^{2}$.
	
	For the matrix $H_Z$, it is straightforward to verify that the number of phase-flip operators (Pauli $Z$) appearing in each stabilizer generator is even. Consequently, the corresponding encoding procedure and the associated quantum circuit do not present any particular difficulty.
	
	However, a more detailed analysis is required for the encoding of the generators associated with the matrix $H_X$.
	
	In $H_X$, for all indices $i$ and $j$ such that $1\leq i \leq 3k^{2}$ and $1\leq j \leq 6k^{2}$, we have:
	
	\begin{itemize}
		\item[1.] If $i\leq 3k$, then $H_X[i,j]=1$ except at the following $k$ positions:
		\begin{itemize}
			\item[i.] when $j=m(3k)+i$ for all $m\in[1,k-1]$ such that $m(3k)+i \leq 3k^{2}$,
			\item[ii.] and when $j=3k^{2}+i$.
		\end{itemize}
		
		\item[2.] If $i\geq 3k$, then $H_X[i,j]=1$ except at the following $k$ positions:
		\begin{itemize}
			\item[i.] when $i=m(3k)+j$ for all $m\in[1,k-1]$ such that $m(3k)+j \leq 3k^{2}$,
			\item[ii.] when $j=m(3k)+i$ for all $m\in[1,k-1]$ such that $m(3k)+i \leq 3k^{2}$,
			\item[iii.] and when $j=3k^{2}+i$.
		\end{itemize}
	\end{itemize}
	
	\begin{example}
		Let us consider the case $k=4$. We obtain:
		
		\begin{itemize}
			\item[1.] If $i\leq 12$, then $H_X[i,j]=1$ except at the positions
			\begin{itemize}
				\item[i.] $j=12+i,\; 24+i,\;36+i$,
				\item[ii.] and $j=48+i$.
			\end{itemize}
			
			\item[2.] If $i\geq 12$, then $H_X[i,j]=1$ except at
			\begin{itemize}
				\item[i.] the positions satisfying $i=j+12,\; i=j+24,\; i=j+36$,
				\item[ii.] the positions $j=12+i,\;24+i,\;36+i$ whenever these indices are less than or equal to $48$,
				\item[iii.] and $j=48+i$.
			\end{itemize}
		\end{itemize}
		
		Hence, the first two generators of the quantum code
		$\mathcal{QL}_4 = [[96,16,4]]$, obtained for $k=4$, are given by
		
		\[
		g_1=\left\{
		\begin{array}{ll}
			I & \text{on all positions except} \\
			X & \text{at positions } j=13,25,37,49,
		\end{array}
		\right.
		\]
		
		\[
		g_2=\left\{
		\begin{array}{ll}
			I & \text{on all positions except} \\
			X & \text{at positions } j=14,26,38,50.
		\end{array}
		\right.
		\]
		
		The thirteenth generator is similarly given by
		
		\[
		g_{13}=\left\{
		\begin{array}{ll}
			I & \text{on all positions except} \\
			X & \text{at positions } j=1,25,37,61.
		\end{array}
		\right.
		\]
		
	\end{example}
	
	\begin{remark}
		We observe that, for the new quantum code $\mathcal{QL}_k$, each $X$-type generator contains exactly $k$ bit-flip operations within its support. Consequently, the matrix $H_X$ takes the following structural form:
	\end{remark}
	
	\begin{equation*}
		H_X=\left[ \begin{array}{cccccccccccccccc|c}
			I&...&I&X&I&...&I&X&I&...&...&I&X&I&...&I&\multirow{15}{*}{$I_{3k^{2}}(X)$}\\
			\vdots&I& &I&X&I&...&I&X&I&...&...&I&X&I&\vdots&\\
			\vdots&& \ddots&&\ddots&\ddots& \ddots&&\ddots&\ddots&\ddots&&&\ddots&\ddots&\vdots&\\
			I&...&...&I&...&I&X&I&...&I&X&I&...&...&I&X&\\
			X&I&...&...&I&...&I&X&I&...&I&X&I&...&...&I&\\
			I&X&I&...&...&I&...&I&X&I&...&I&X&I&...&I&\\
			\vdots&\ddots& \ddots&\ddots&&& \ddots&&\ddots&\ddots&\ddots&&\ddots&\ddots&\ddots&\vdots&\\
			I&...&I&X&I&...&...&I&...&I&X&I&...&I&X&I&\\
			X&I&...&I&X&I&...&...&I&...&I&X&I&...&I&\\
			I&X&I&...&I&X&I&...&...&I&...&I&X&I&...&I&\\
			\vdots&\ddots& \ddots&\ddots&&\ddots& \ddots&\ddots&&&\ddots&&\ddots&\ddots&\ddots&\vdots&\\
			I&...&I&X&I&...&I&X&I&...&...&I&...&I&X&I&\\	
			X&I&...&I&X&I&...&I&X&I&...&...&I&...&I&X&\\
			I&X&I&...&I&X&I&...&I&X&I&...&...&I&...&I&\\
			\vdots&\ddots&\ddots&\ddots&&\ddots& \ddots&\ddots&&\ddots&\ddots&\ddots&&&\ddots&\vdots&\\
		\end{array}\right] 
	\end{equation*}
	where the matrix $I_{3k^{2}}(X)$ is defined by
	\[
	I_{i,j}(X)=
	\begin{cases}
		I & \text{if } i\neq j,\\
		X & \text{if } i=j.
	\end{cases}
	\]

	\subsection{Quantum encoding circuit}
	
	For the encoding of the new quantum code $\mathcal{QL}_k$, the input register has length $k^{2}$ qubits, while the output register contains $6k^{2}$ qubits. Therefore, $5k^{2}$ ancilla qubits initialized in the state $|0\rangle$ are required.
	
	The encoding circuit corresponding to the elements of $H_X$ is constructed as follows:
	
	\begin{itemize}
		\item[1.] For every circuit row such that $i\leq 3k$:
		\begin{itemize}
			\item[-] The root qubit is chosen as the $(3k+i)$-th qubit, denoted $q_{3k+i}$.
			\item[-] A Hadamard gate $H$ is applied to $q_{3k+i}$ to generate a superposition.
			\item[-] A sequence of CNOT gates controlled by $q_{3k+i}$ is then applied:
			
			CNOT$(3k+i \rightarrow 2\times(3k)+i)$,
			
			CNOT$(3k+i \rightarrow 3\times(3k)+i)$,
			
			$\vdots$
			
			CNOT$(3k+i \rightarrow (k-1)\times(3k)+i)$,
			
			and finally
			
			CNOT$(3k+i \rightarrow 3k^{2}+i)$.
		\end{itemize}
		
		\item[2.] For every circuit row such that $i\geq 3k$:
		\begin{itemize}
			\item[-] The root qubit is chosen as the $(i-3k)$-th qubit, denoted $q_{i-3k}$.
			\item[-] A Hadamard gate is applied to $q_{i-3k}$.
			\item[-] Controlled CNOT gates are then applied:
			
			CNOT$(i-3k \rightarrow 2\times(3k)+i)$,
			
			CNOT$(i-3k \rightarrow 3\times(3k)+i)$,
			
			$\vdots$
			
			CNOT$(i-3k \rightarrow m\times(3k)+i)$ with $m\times(3k)+i \leq 3k^{2}$,
			
			and finally
			
			CNOT$(i-3k \rightarrow 3k^{2}+i)$.
		\end{itemize}
	\end{itemize}
	
	\begin{remark}
		As in any quantum encoding scheme, if the root qubit is initialized in the state $|1\rangle$, an $X$ gate is applied prior to the Hadamard gate.
	\end{remark}

	\subsection{Simplified Decoding Procedure}
	Thanks to the highly regular structure of the quantum encoding for the code $\mathcal{QL}_k$, the decoding process can be significantly simplified.  
	
	Each $X$-type stabilizer acts on exactly $k$ qubits within its support, while the $Z$-type stabilizers of $H_Z$ exhibit even parity across each generator. This structure allows for a CSS-type decoding to be implemented in two sequential stages:  
	
	\begin{enumerate}
		\item [-] Phase-flip error correction : $Z$-type errors can be efficiently detected and corrected using the $X$-type stabilizers derived from $H_X$.
		\item [-]Bit-flip error correction ($X$-errors): $X$-type errors can be detected and corrected via the $Z$-type stabilizers from $H_Z$.
	\end{enumerate}
	
	Furthermore, the repetitive architecture of the encoding circuit enables local block-wise correction. Specifically, each root qubit together with its associated $k$ target qubits forms a block where syndromes can be measured and corrected in parallel. This block-based structure substantially reduces both the number of measurements and the required quantum operations during decoding, thereby enhancing the overall efficiency of the procedure.

	\section{Conclusion}
	
	We have introduced a novel family of quantum codes constructed from two families of classical linear codes, $L_k$ and $L_k^+$. The resulting CSS construction defines stabilizer codes with parameters
	$$QL_k = [[6k^2, k^2, d]],$$
	with a minimum distance $d = 4$ for $k \geq 4$. The code size grows quadratically with $k$ while maintaining a constant rate of $1/6$, providing a favorable trade-off between the number of logical qubits and physical qubits.
	
	This family is distinguished by its rich combinatorial structure, inherited from the incidence and orthogonality relations of the underlying classical codes. The stabilizer matrices are sparse, endowing the codes with an LDPC character and strong locality, where each qubit participates in a bounded number of stabilizers. This combinatorial regularity facilitates theoretical analysis, the design of efficient decoding algorithms, and implementation on realistic quantum architectures.
	
	These results open multiple perspectives: improving the minimum distance while preserving LDPC structure and a constant rate, generalizing the construction to multiple or asymmetric products of classical codes, and developing new decoding algorithms that leverage the combinatorial regularity. Exploration of these directions could lead to more robust quantum codes suitable for large-scale applications.
	
	\textbf{Acknowledgments.} The author thanks colleagues and collaborators for valuable discussions and feedback that contributed to this work.

	


	\vspace{3mm}
	\begin{center}
		\begin{parbox}{118mm}{\footnotesize
				
				Luc RABEFIHAVANANA
				
				ORCID: \texttt{https://orcid.org/0009-0000-4977-8380}
				
				Department of Mathematics and computer science
				
				University of Antananarivo-Madagascar
				
				E--mail: \texttt{lucrabefihavanana@gmail.com}
				
				\vspace{2mm}
				Harinaivo ANDRIATAHINY
				
				Department of Mathematics and computer science
				
				University of Antananarivo-Madagascar
				
				E--mail: \texttt{hariandriatahiny@gmail.com }
				
				\vspace{2mm}
				
				Ferdinand RANDRIAMIARAMPANAHY
				
				Department of Mathematics and computer science
				
				University of Antananarivo-Madagascar
				
				E--mail: \texttt{randriamiferdinand@gmail.com}
				
			}
			
		\end{parbox}
	\end{center}
	
	\end{document}